\documentclass[aps,prb,reprint,amsmath,amssymb,superscriptaddress,showpacs]{revtex4-1}
\usepackage{graphicx}
\usepackage{color}
\usepackage{hyperref}
\usepackage{multirow}

\newcommand{\tcur}{T^{\rm{Curie}}}
\newcommand{\tc}{T_{\rm C}}
\newcommand{\mB}{\mu_{\rm B}}

\begin{document}

\title{Depth profile of the ferromagnetic order in a $\rm YBa_2Cu_3O_7/La_{2/3}Ca_{1/3}MnO_3$ superlattice on a LSAT substrate: a polarized neutron reflectometry study}

\author{M.~A.~\surname{Uribe-Laverde}}
\email{miguelangel.uribelaverde@unifr.ch}
\author{D.~K.~Satapathy}
\altaffiliation{Present address: Department of Physics, Indian Institute of Technology Madras, 600036 Chennai, India}
\author{I.~Marozau}
\author{V.~K.~Malik}
\author{S.~Das}
\author{K.~Sen}
\affiliation{University of Fribourg, Department of Physics and Fribourg Centre for Nanomaterials, Chemin du Mus\'ee 3, CH-1700 Fribourg, Switzerland}

\author{J.~Stahn}
\affiliation{Laboratory for Neutron Scattering Paul Scherrer Institut, CH-5232 Villigen, Switzerland}
\author{A.~R\"{u}hm}
\affiliation{Max-Planck-Institut f\"ur Intelligente Systeme, Heisenbergstrasse 3, D-70569 Stuttgart, Germany}

\author{J.~-H.~Kim}
\author{T.~Keller}
\affiliation{Max-Planck-Institut f\"ur Festk\"orperforschung, Heisenbergstrasse 1, D-70569 Stuttgart, Germany}

\author{A.~Devishvili}
\author{B.~P.~Toperverg}
\affiliation{Institute of Solid State Physics, Ruhr-Universit\"at Bochum,  D-44801 Bochum, Germany}

\author{C.~Bernhard}
\affiliation{University of Fribourg, Department of Physics and Fribourg Centre for Nanomaterials, Chemin du Mus\'ee 3, CH-1700 Fribourg, Switzerland}


\date{\today}

\begin{abstract}
Using polarized neutron reflectometry (PNR) we have investigated a [$\rm YBa_2Cu_3O_7$(10\,nm)\,/\,$\rm La_{2/3}Ca_{1/3}MnO_3$(9\,nm)]$_{10}$ (YBCO/LCMO) superlattice grown by pulsed laser deposition on a $\rm{La_{0.3}Sr_{0.7}Al_{0.65}Ta_{0.35}O_{3}}$ (LSAT) substrate. Due to the high structural quality of the superlattice and the substrate, the specular reflectivity signal extends with a high signal-to-background ratio beyond the fourth order superlattice Bragg peak. This allows us to obtain more detailed and reliable information about the magnetic depth profile than in previous PNR studies on similar superlattices that were partially impeded by problems related to the low temperature structural transitions of the SrTiO$_3$ substrates~[J. Stahn \emph{et al.}, Phys. Rev. B \textbf{71}, 140509 (2005)]. In agreement with the previous reports, our PNR data reveal a strong magnetic proximity effect showing that the depth profile of the magnetic potential differs significantly from the one of the nuclear potential that is given by the YBCO and LCMO layer thickness. We present fits of the PNR data using different simple block-like models for which either a ferromagnetic moment is induced on the YBCO side of the interfaces or the ferromagnetic order is suppressed on the LCMO side. We show that a good agreement with the PNR data and with the average magnetization as obtained from dc magnetization data can only be obtained with the latter model where a so-called depleted layer with a strongly suppressed ferromagnetic moment develops on the LCMO side of the interfaces. The models with an induced ferromagnetic moment on the YBCO side fail to reproduce the details of the higher order superlattice Bragg peaks and yield a wrong magnitude of the average magnetization. We also show that the PNR data are still consistent with the small, ferromagnetic Cu moment of $0.25\,\mB$ that was previously identified with  x-ray magnetic circular dichroism and x-ray resonant magnetic reflectometry measurements on the same superlattice~[D.K. Satapathy \emph{et al.}, Phys. Rev. Lett. \textbf{108}, 197201 (2012)]. We emphasize that the ferromagnetic moment of these Cu moments is apparently induced by the coupling to the Mn moments suggesting that the depleted layer cannot be a so-called \textquotedblleft dead\textquotedblright\ layer that is entirely not magnetic. The observed strong suppression of the ferromagnetic moment in the depleted layer thus may be related to a canted antiferromagnetic or an oscillatory type of ordering of the Mn moments that is not detected with the PNR technique.
\end{abstract}


\maketitle

The interaction of the competing superconducting (SC) and ferromagnetic (FM) order parameters is a fascinating topic that has been extensively studied theoretically and experimentally \cite{Bulaevskii_1985,Buzdin_2005,Bergeret_2005}. In recent years significant progress has been made with thin-film heterostructures from conventional superconductors and elemental or alloy ferromagnets where effects such as domain wall superconductivity~\cite{Buzdin_2003}, critical temperature oscillations with the thickness of the FM layer in SC/FM/SC junctions~\cite{Kehrle_2012} and a long-range proximity effect of a spin-triplet SC state through composite FM layers with a non-collinear magnetic order~\cite{Volkov_2003}, were predicted theoretically and confirmed experimentally~\cite{Kontos_2002,Blum_2012,Khaire_2010,Robinson_2010,Yang_2004}.  These developments have inspired concepts for a new kinds of spintronic devices and applications~\cite{Eschrig_2011,Pugach_2012}.\\\\
Driven by curiosity and encouraged by the potential for applications, researchers have also started to investigate heterostructures involving high-$\tc$ cuprate superconductors (HTSC) and ferromagnetic manganites~\cite{Przyslupski_1997,Goldman_2001,Habermeier_2001,Sefrioui_2003,Holden_2004,Werner_2010}. The common perovskite-related structure and similar in-plane lattice parameters, together with recent improvements in film deposition techniques, enable the layer by layer controlled epitaxial growth of multilayers and superlattices (SLs) with very sharp interfaces~\cite{Sefrioui_2003,Malik_2012}. Experiments on this kind  of oxide SC/FM heterostructures revealed effects such as a change in the SC critical temperature, $\tc$, related to the presence and thickness of the FM layers, and vice-versa~\cite{Sefrioui_2003,Werner_2010}; a SC related giant magneto-resistance in FM/SC/FM trilayers~\cite{Pena_2005}; the enhancement of $\tc$ by an external magnetic field~\cite{Nemes_2008,Dybko_2009} and even a SC induced modulation of the magnetic moment in the FM layers~\cite{Hoppler_2009}. These observations provide encouraging evidence for a sizable interaction between the SC and FM order parameters in these cuprate/manganite multilayer systems. They also show that not only the superconducting but also the magnetic properties of these oxide heterostructures are extremely versatile and need to be thoroughly investigated.\\\\
Polarized neutron reflectometry (PNR) measurements on $\rm{YBa_2Cu_3O_7}$/$\rm{La_{2/3}Ca_{1/3}MnO_3}$ (YBCO/LCMO) SLs have indeed revealed an unusual kind of magnetic proximity effect that gives rise to a significant change of the FM order in the vicinity of the YBCO/LCMO interfaces. PNR measurements on SLs with equally thick YBCO and LCMO layers, for which the even order superlattice Bragg-peaks should be absent for symmetry reasons, showed that a 2nd order Bragg peak appears and gains considerable intensity below the FM transition, $\tcur$, of the LCMO layers ~\cite{Stahn_2005}. This observation clearly showed that the depth profile of the magnetic potential has a lower local symmetry than the nuclear one. In other words, the FM moment is either significantly modified (reduced) on the LCMO side of the interface or a FM moment is induced on the YBCO side. Two different models were proposed in Ref.~\onlinecite{Stahn_2005} to describe this magnetic proximity effect: a so-called \textquotedblleft dead layer model\textquotedblright\- with a strongly reduced magnetization on the LCMO side, and a so-called \textquotedblleft inverse proximity effect model\textquotedblright\-  where ferromagnetic Cu moments antiparallel to the Mn moments in LCMO are induced on the YBCO side of the interface. It was not possible to distinguish between these two possibilities since the PNR data contained only a fairly limited range in momentum space such that the superlattice Bragg peaks beyond the 2nd order were not observed. Later on, it was found that this is due to a problem of the SrTiO$_3$ (STO) substrates for which a structural transition below 100 K can give rise to a buckling of the surface with micrometer-sized structural domains that are tilted up to 0.5 degree~\cite{Hoppler_2008,Hoppler_2009}. It was shown that this tilting extends into the SL on top of the STO substrate and thus gives rise to a large broadening or even a well pronounced splitting of the PNR specular signal along the 2$\theta$ direction of the detector.\\\\
In the meantime, i.e. before this problem with the STO substrates had been fully appreciated, additional experiments of YBCO/LCMO SLs on STO substrates were reported that supported either the dead layer model or the inverse proximity effect model. Hoffmann et al. reported an enhancement in the Mn~3$d$ occupation next to the interface and concluded that their PNR data support the dead layer model~\cite{Hoffmann_2005}. On the other hand, Chakhalian et al. reported  x-ray magnetic circular dichroism (XMCD) data which established the presence of a ferromagnetic Cu moment and thus were interpreted in terms of the \textquotedblleft inverse proximity effect model\textquotedblright\cite{Chakhalian_2006}. Seemingly contradictory results were also reported concerning theoretical calculations. The density functional theory calculations performed by Luo et al. predicted a strongly reduced ferromagnetic or even antiferromagnetic coupling for the Mn atoms next to the interfaces~\cite{Luo_2008}, whereas the theoretical work of Salafranca et al. concluded that the negative spin polarization of Cu in YBCO layers is a key ingredient to explain the observed enhancement of $\tc$ with magnetic field observed in LCMO/YBCO/LCMO trilayers~\cite{Salafranca_2010,Nemes_2008}.\\\\
More recently, some of the present authors have been shedding new light on this seemingly contradictory issue of the magnetic proximity effect. They investigated a YBCO/LCMO SL on a $\rm{La_{0.3}Sr_{0.7}Al_{0.65}Ta_{0.35}O_{3}}$ (LSAT) substrate which is well lattice-matched and avoids the complications of the buckling of the STO substrates \cite{Satapathy_2012}. On this SL they performed a combination of PNR, XMCD and x-ray resonant magnetic reflectometry (XRMR) which showed that the magnetic proximity effect at the YBCO/LCMO interface involves in fact both effects, i.e. a suppression of the FM moment on the LCMO side and, yet, an induced ferromagnetic Cu moment on the YBCO side. The latter was firmly established by the XMCD and especially by the XRMR measurements at the Cu-L edge which confirmed that the ferromagnetic Cu moments reside within the YBCO layers. Specifically, they demonstrated that the ferromagnetic Cu moments do not arise from a small amount of Cu ions that might have been incorporated in the LCMO layers. The existence of a layer with a depleted FM moment on the LCMO side was inferred from the analysis of the PNR data which extend with a sufficient signal-to-noise-ratio to the 4th order superlattice Bragg peak. In this manuscript we present the details of the analysis of these PNR data which could not be shown in the previous short letter. Specifically, we show that the \textquotedblleft dead layer model\textquotedblright, or rather the \textquotedblleft depleted layer model\textquotedblright\- as we prefer to call it for the reasons given below, accounts very well for the PNR data. We also show that alternative models fail to reproduce important features of the PNR curves.\\\\
We remark that this observation has important implications. In combination with the XMCD and XRMR data, which reveal the existence of an induced FM Cu moment on the YBCO side of the interface, it suggests that a strongly reduced but finite FM order or possibly even a non-collinear magnetic order of the Mn moments persists in the depleted layer which mediates the antiparallel coupling between the Cu moments in YBCO and the Mn moments in the central part of the LCMO layers. The so-called depleted layer therefore should not be considered as a magnetically \textquotedblleft dead layer\textquotedblright, instead it seems very much alive and may play an important role, for example, in the recently reported long-range SC proximity effect\cite{Visani_2012,Golod_2012}.
\section{Experimental}
The [YBCO (10nm)  /LCMO (9nm)]$_{10}$ SL was grown on a LSAT substrate using the pulsed laser deposition technique (PLD) as described elsewhere~\cite{Malik_2012}. The substrate with a size of 10$\times$10\,mm$^2$ and a thickness of 0.5\,mm was purchased from Crystec. The monolayer by monolayer growth has been controlled with \textit{in situ} reflection high-energy electron diffraction (RHEED). Resistivity measurements were performed using the four-point probe option of a physical properties measurement system (PPMS) from Quantum Design (Model QD6000). The magnetization was measured on a small piece cut from the corner of the sample using the vibrating sample magnetometer (Model P525) option of the PPMS system. X-ray reflectivity (XRR) measurements were carried out with the UE56/2-PGM1 beamline at BESSY, using the MPI-IS ErNST endstation. The PNR measurements were performed with the two-axis reflectometers NREX at FRM-II in Munich, D, and SuperADAM at ILL in Grenoble, F. Magnetic fields up to 4\,kOe oriented perpendicular to the scattering plane and parallel to the film surface were produced with electromagnets. The temperatures and applied magnetic fields were as follows: (4\,K, 4000\,Oe, NREX), (100\,K, 4000\,Oe, NREX), (300\,K, 0\,Oe, NREX), (10\,K, 100\,Oe, SuperADAM), (100\,K, 100\,Oe, SuperADAM), (300\,K, 0\,Oe, SuperADAM).\\\\
The PNR curves have been fitted using the SUPERFIT package originally developed at the Max-Planck-Institut Stuttgart~\cite{Superfit}. This package uses the maximum likelihood probability algorithm to define the minimizing function, the minimization is performed with the MINUIT package~\cite{James_1975}. The likelihood estimator is defined as
\[l=\sum_{i=1}^N\left[y_i-x_i+x_i\ln\left(\frac{x_i}{y_i}\right)\right],\]
where $N$ is the total number of data points, $x_i$ is the measured intensity and $y_i$ is the calculated intensity using the super-matrix formalism~\cite{Ruehm_1999}. Reported in the following is the reduced likelihood estimator $l^{ \rm{red}}=l/(N-n)$, where $n$ is the number of fitted parameters.
We used a modified version of the SUPERFIT program, which allows us to fit several datasets simultaneously with global parameters, to fit more reliably the common structural parameters.\\\\
The simulations of the XRR data have been performed using the software package GenX~\cite{Bjorck_2007}.
\section{Results and Discussion}
\begin{figure}[t!]
\centering
\includegraphics[width=8.6cm]{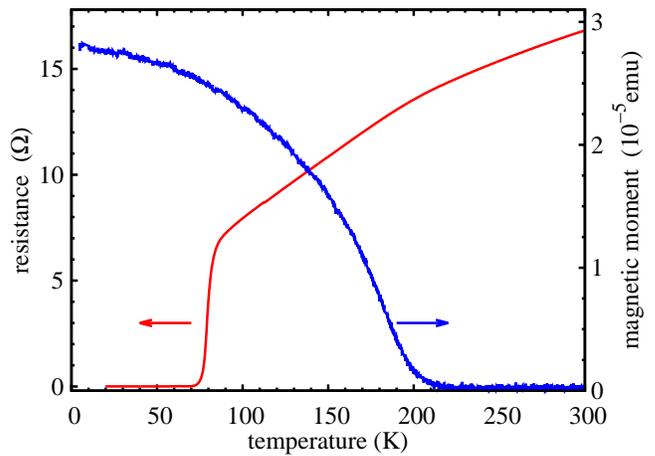}
\caption{Temperature dependence of the resistance and the field-cooled (FC) magnetization at an applied field of $H_{\rm{app}}=100$\,Oe as measured on the YBCO/LCMO superlattice. It shows the onset of the superconducting transition at $\tc=88$\,K and the ferromagnetic transition at $\tcur=201$\,K.\label{RT}}
\end{figure}
\begin{figure}[t!]
\centering
\includegraphics[width=8.6cm]{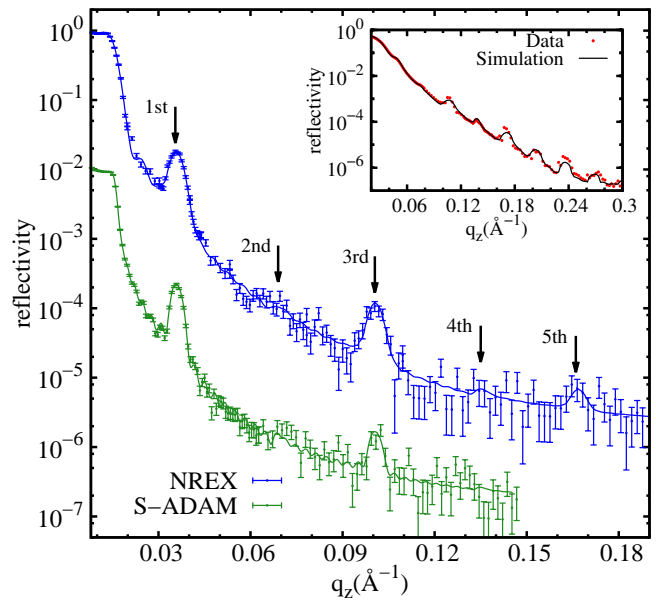}
\caption{Unpolarized neutron reflectivity curves of the YBCO/LCMO SL measured at room temperature with the NREX and SuperADAM instruments. The curves are vertically shifted for clarity. Symbols show the experimental data and solid lines the best fits that were obtained by fitting both curves simultaneously. The arrows mark the position of the SLBPs. Inset: Symbols show the non-resonant x-ray reflectivity curve at 300K. The solid line shows a simulation using the parameters as obtained from the fits of the neutron reflectometry curves.\label{PNR_RT}}
\end{figure}
\begin{figure}[t]
\centering
\includegraphics[width=8.6cm]{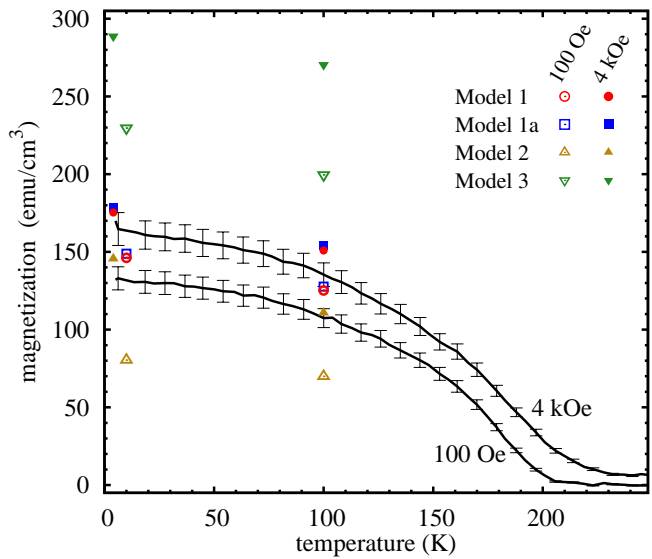}
\caption{Comparison of the average magnetic moment as determined experimentally from field-cooled dc magnetization measurements at 100\,Oe and 4\,kOe (solid lines) and calculated from the magnetic potential obtained with model 1 (red circles), model 1a (blue squares), model 2 (golden triangles) and model 3 (green triangles) from the fits to the PNR curves measured at 100\,Oe (open symbols) and 4\,kOe (solid symbols). The size of the symbols of the calculated magnetic moments reflects the error bars. The error bars of the dc magnetization data arise from the statistical errors and the uncertainty of the volume of the small piece used for the dc magnetization measurements.\label{FC}}
\end{figure}

Figure~\ref{RT} shows the temperature dependence of the resistance and the field-cooled (FC) magnetization of the YBCO/LCMO SL. The resistance shows a sharp decrease at the onset of the superconducting transition at $\tc=88$\,K and vanishes below 72\,K. The ferromagnetic transition at $\tcur= 201$\,K is evident from the sudden increase in the magnetic moment as well from a kink in the temperature dependence of the normal state resistance. The latter feature originates from the insulator-to-metal transition in the LCMO layer which coincides with the ferromagnetic ordering.\\\\
Figure~\ref{PNR_RT} shows the unpolarized neutron reflectivity curves that were obtained at room temperature. The slight differences between the two curves are due to the different experimental configurations, for SuperADAM we used a high resolution setup whereas for NREX the signal-to-noise ratio was enhanced at the expense of a lower resolution. Both curves exhibit a sharp reflection edge and a set of well-defined superlattice Bragg peaks (SLBP). The latter originate from the constructive interference between the neutron waves that are reflected from all the interfaces of the SL. The position, width and intensity of the SLBPs contain the information about the average value and the variation of the thickness of the individual YBCO and LCMO layers. Additional information about the roughness of the SL is contained in the form of the overall decrease toward large momentum transfer of the reflectivity curve beyond the reflection edge. In the absence of roughness this decrease follows a $q_z^{-4}$ law, where $q_z$ is the value of the normal momentum transfer. The roughness makes this overall decrease of the intensity of the reflectivity curve even faster.\\\\
The high resolution data from SuperADAM also show a high frequency oscillation in the $q_z$ range between the reflection edge and the 1st SLBP. These are so-called Kiessig fringes that originate from the interference between the reflections from the surface (ambient/LCMO) and the bottom (interface with the LSAT substrate) of the SL. These features testify for the high quality of our SL, from their period we can deduce the thickness of the entire SL. The thickness of the YBCO/LCMO bilayers can be inferred from the position of the SLBPs. The information about the thickness ratio of the YBCO and LCMO layers is contained in the intensity variation of the even and odd order SLBPs. For example, the even order SLBPs are entirely suppressed if the YBCO and LCMO layers have exactly the same thickness. This is a destructive interference phenomenon that originates from a $\pi$-phase shift between the neutron waves that are reflected at the YBCO/LCMO and the LCMO/YBCO interfaces. It arises because the scattering potential exhibits a step-like increase at one of the interfaces and a corresponding decrease on the other one. In the reflectivity curves in Fig.~\ref{PNR_RT} the intensity of the 2nd and 4th order SLBPs is indeed much weaker than the one of the 1st and 3rd order SLBPs. The finding that the suppression of the even order SLBP intensity is not complete, i.e. that a small increase is observed at the position of the 2nd and the 4th order SLBPs, shows that there is a small mismatch between the thickness of YBCO and LCMO layers.\\\\
As a starting point of our analysis we have simultaneously fitted the two room temperature unpolarized neutron reflectivity curves to extract the structural parameters. The result of the best fit is shown by the solid lines in Fig.~\ref{PNR_RT} and can be seen to describe the experimental data (symbols) very well. The obtained values of the nuclear scattering length density (SLD) are $\rho^{\rm N}_{\rm YBCO}=4.511(12)\times 10^{14}$\,m$^{-2}$ and $\rho^{\rm N}_{\rm LCMO}=3.531(12)\times 10^{14}$\,m$^{-2}$, the thickness parameters are $d_{\rm YBCO}=9.773(72)$\,nm and $d_{\rm LCMO}=9.087(72)$\,nm. The latter correspond to approximately 8 and 23 unit cells per YBCO and LCMO layer, respectively. The obtained roughness of the film of 8.5(2)\,\AA~is similar to the size of a YBCO unit cell and testifies for the high quality of the SL. As a consistency check, we have used the obtained parameters for the thickness and roughness to simulate a X-ray reflectivity curve that was measured on the same SL. The inset of Fig.~\ref{PNR_RT} shows the good agreement with the experimental curve (symbols) and the calculation (solid lines) for which only the value of the SLD has been adjusted. \\\\
\begin{figure*}[t!]
\centering
\includegraphics[width=15cm]{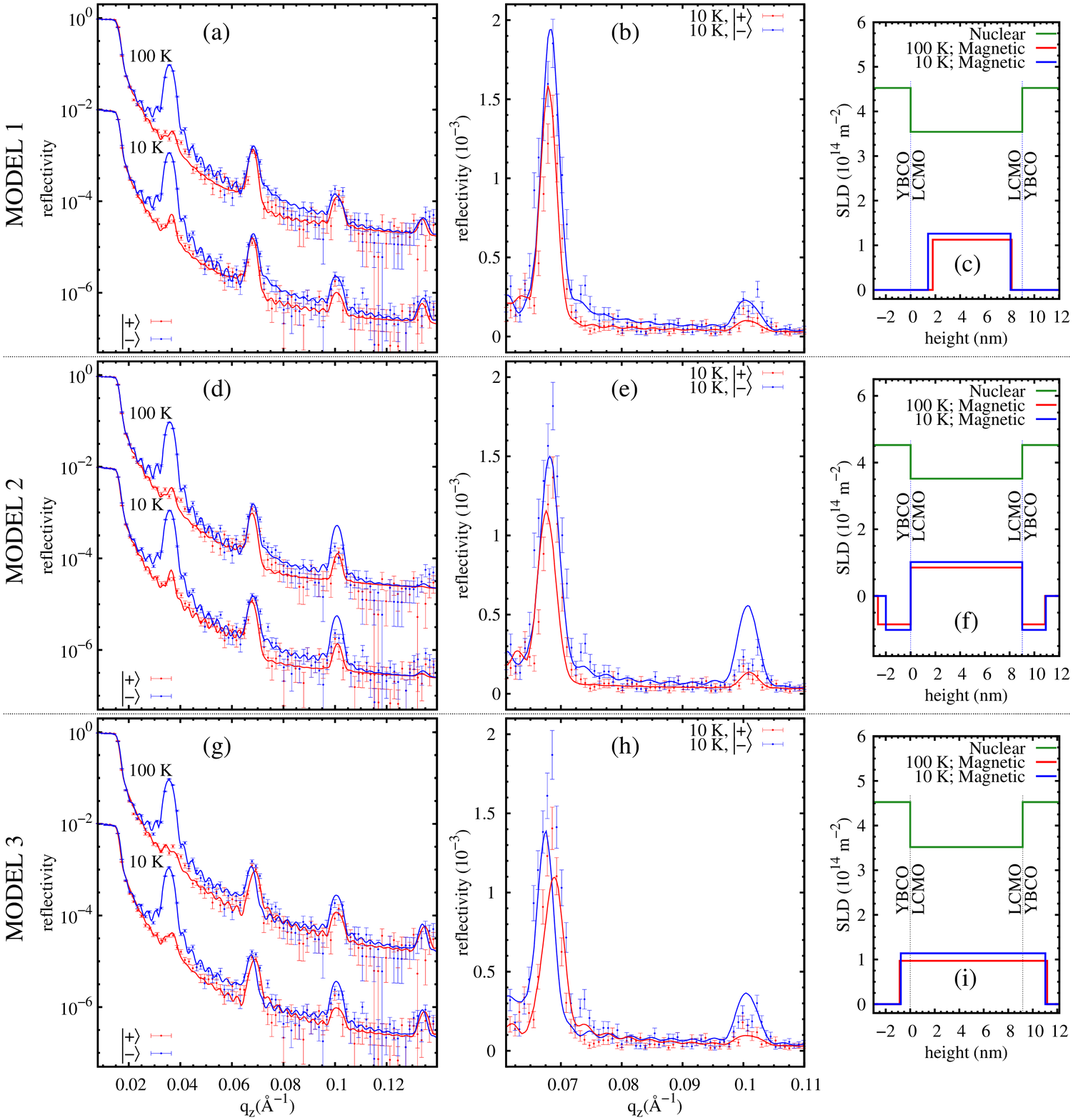}
\caption{Left: (a) Polarized reflectivity curves of the YBCO/LCMO SL measured at low temperature at SuperADAM for spin-up $|+\rangle$ and spin down $|-\rangle$ polarization of the neutron spin with respect to the direction of the applied magnetic field of 100\,Oe. For clarity the curves at 10K are vertically shifted down by a factor of $10^{2}$. (b) Close-up on a linear intensity scale in the region of the 2nd and 3rd SLBP to aid the comparison with the best fit of the calculations for the depleted layer model (solid lines). The depth profiles of the used nuclear and magnetic scattering length densities are shown in (c). The same data are shown in (d)-(f) together with the best fit using the model of an inverse magnetic proximity effect and, in (g)-(i), for the model of an induced FM moment in YBCO that is parallel to the one in LCMO.\label{SADAM}}
\end{figure*}
\begin{figure*}[t!]
\centering
\includegraphics[width=15cm]{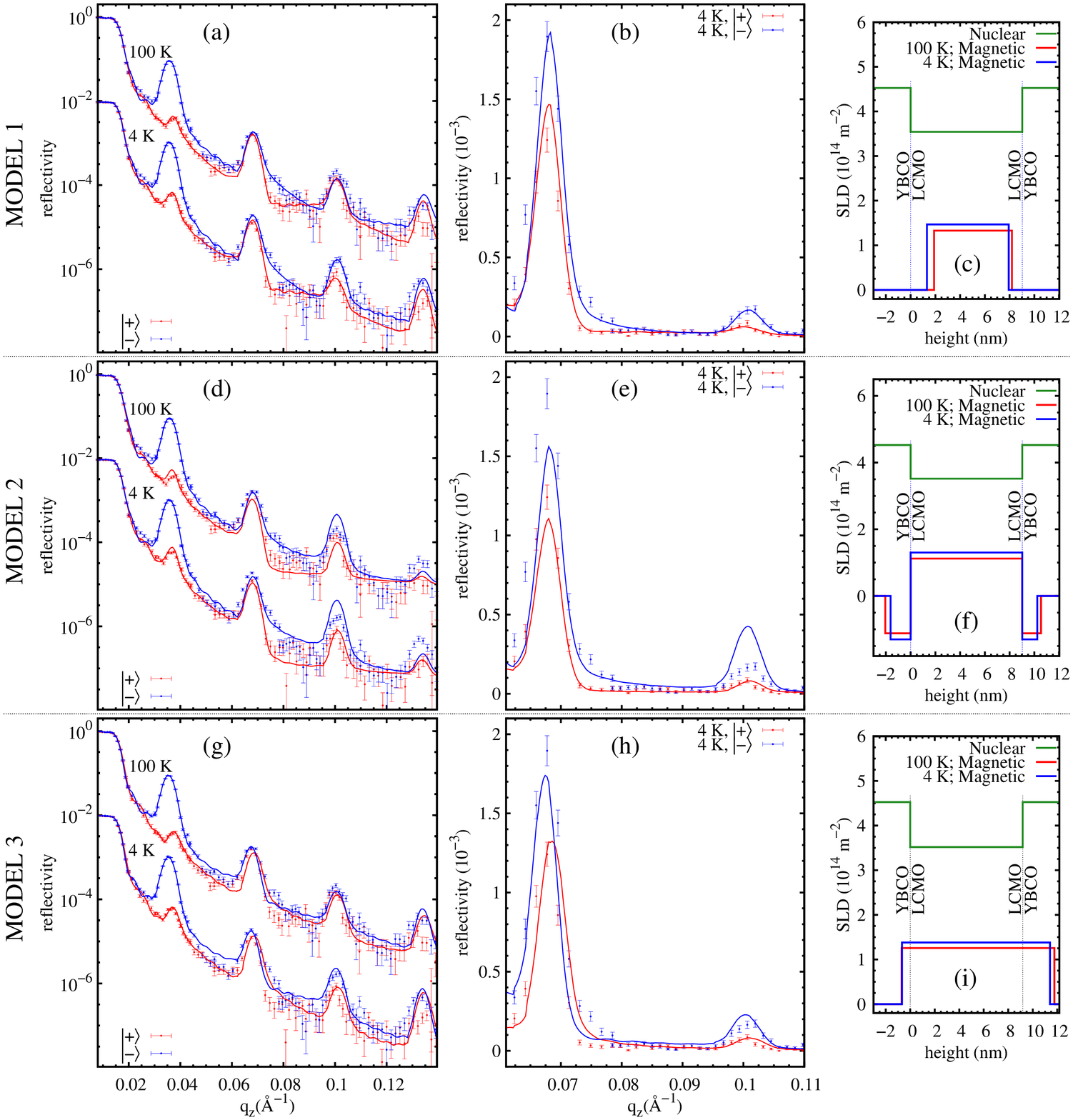}
\caption{Same as in Fig.~\ref{SADAM} but for the PNR data measured at NREX with an applied field of 4\,kOe.\label{NREX}}
\end{figure*}
\begin{figure*}[t!]
\centering
\includegraphics[width=15cm]{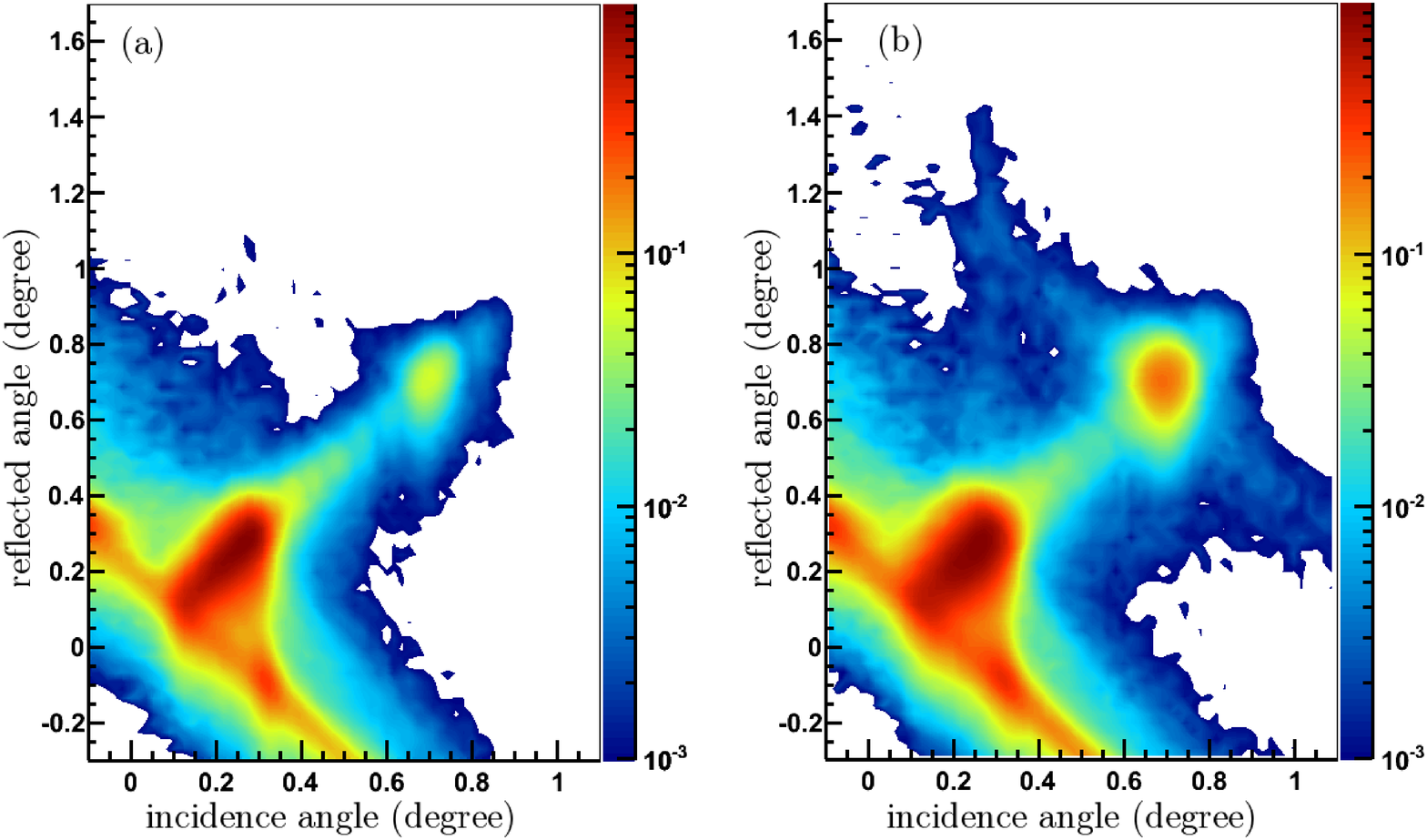}
\caption{Maps of the off-specular reflection of the YBCO/LCMO superlattice measured a) with unpolarized neutrons at 300\,K and b) for the $|+\rangle$ spin channel at 4\,K after field-cooling in a field of 100\,Oe.\label{offspec}}
\end{figure*}
Figures~\ref{SADAM} and~\ref{NREX} show the low temperature PNR curves which have been measured at an applied magnetic field of 100\,Oe and 4\,kOe with SuperADAM and NREX, respectively. In fitting these PNR curves the structural parameters as obtained from the room temperature curves (as described above) have been used as a constraint that can vary only within the error bar. This seems well justified, since the expected changes due to the temperature dependence of the lattice parameters of YBCO and LCMO are well within these error bars. We have also confirmed with temperature dependent X-ray measurements on a LSAT substrate that it does not undergo any anomalous structural change. In addition, Figure~\ref{offspec} shows the off-specular scattering at 300 and 4\,K which confirm that the anomalous broadening and splitting of the specular reflection curves that was previously reported for similar SLs on STO substrates at temperatures below 100\,K\cite{Hoppler_2008,Hoppler_2009,Hoppler_2010} is entirely absent for the present YBCO/LCMO SL on LSAT. The comparison of the maps in Figs.~\ref{offspec}a and \ref{offspec}b shows that for the reflection edge and the SLBPs the width in the off-specular direction is small and does not change significantly between 300 and 4\,K. The quality of the specular reflectivity curves at low temperature accordingly remains very high. As shown in Figs.~\ref{SADAM} and~\ref{NREX}, the intensity of the reflectivity curves does not fall off significantly faster at 10\,K or 4\,K than at 300\,K and the signal remains well above the background level for the $q_z$ values up to 0.14\,\AA$^{-1}$ which includes the 4th order SLBP.\\
The shape of the reflectivity curves below $\tcur=201$\,K, as shown in Figs~\ref{SADAM} and~\ref{NREX}, depends on whether the incident neutron spin is parallel ($|+\rangle$) or antiparallel ($|-\rangle$) to the applied magnetic field. This is due to the additional magnetic potential which is comparable in size to the nuclear one and for which the sign depends on the orientation of the FM moments with respect to the one of the neutron spins. It is also evident that the 2nd and the 4th order SLBPs, which were almost absent in the curves at 300\,K, have gained a lot of intensity and became very pronounced features in the PNR curves at 100\,K, 10\,K and 4\,K. As was already discussed in the introduction, this is a clear indication that the magnetic potential has a lower local symmetry than the nuclear one. For the latter the even order SLBPs were almost absent due to the similar thickness of the YBCO and LCMO layers. The magnetic potential due to the ferromagnetic order, does obviously not exhibit a step-like increase right at the YBCO/LCMO interface or a corresponding decrease at the LCMO/YBCO interfaces. Instead, there must be either a significant decrease of the FM moment on the LCMO side of the interface or a corresponding increase on the YBCO side. \\\\
Due to the lack of the phase information in the reflectometry measurement the shape of the depth profile of the magnetic potential cannot be directly extracted from the PNR data. The reflectivity curves can still be analyzed with different realistic models and their validity can be judged based on how well they reproduce the data. As was the case in Ref.~\onlinecite{Stahn_2005}, it may still happen that different models lead to similar results and therefore cannot be distinguished. As shown below, this ambiguity might be overcome with reflectivity curves that extend up to larger $q_z$ values where the differences between the various models become more pronounced.\\\\
In the following we consider three different models and evaluate how well they can reproduce the PNR data. Model 1 corresponds to the depleted layer model that has been outlined in the introduction. For simplicity, a layer with a completely suppressed FM moment is introduced here on the LCMO side of each interface. The thickness of this layer is a fitting parameter.\\\\
Model 2 describes the inverse magnetic proximity effect, where FM moments antiparallel to the Mn moments in LCMO are induced on the YBCO side of the interface. In the first place, one expects that these are the Cu moments which have been observed with the XMCD and XRMR measurements. However, the magnitude of these ferromagnetic Cu moments is reported to be only~$\sim 0.25\,\mB$. Therefore, it remains to be seen whether they can account for the observed large increase of the intensity of the even order Bragg peaks. On the other hand, we cannot completely exclude the possibility that an additional, possibly even larger contribution arises from some Mn ions that may have been incorporated in the YBCO layers, for example due to an unwanted contamination or a diffusion across the interface during the growth.\\\\
Model 3 accounts for a similar case where the induced FM moment on the YBCO side is parallel to the one of the Mn moments in LCMO and to the applied magnetic field. Such a contribution would have to arise solely from the Mn moments, since the XMCD and XRMR data clearly established the antiparallel orientation of the Cu moments.\\\\
For all three models we had to use a modified magnetic potential for the topmost LCMO layer, i.e., a 2-3\,nm thick non-magnetic layer has been introduced at the film surface. This was necessary to reproduce the sizable differences between the $|+\rangle$ and $|-\rangle$ curves in the region right before and after the 1st order SLBP. We suspect that this effect arises from the interaction of the surface layer with the ambient which degrades the FM order in the first few LCMO monolayers.\\\\
In the following we use the quality of the best fits as the criterion to determine the validity of each model. In addition, we compare the average value of the magnetization obtained from the model with the experimental result as measured with dc magnetization.\\\\
Figs.~\ref{SADAM} and~\ref{NREX} show the PNR curves at 100\,K, 10\,K and 4\,K with the best fits for each model (solid lines) together with the obtained depth profiles of the nuclear and magnetic scattering length densities. As was already mentioned, the structural parameters have been constrained to lie within the error bars of the parameters derived from the unpolarized room temperature curves. The obtained reduced likelihood estimators for models 1, 2 and 3 of $l^{\rm red}_{1}=4.12$, $l^{\rm red}_2=5.18$ and $l^{\rm red}_3=7.37$, respectively, are in favor of model 1.
The specific features where models 2 and 3 fail to describe the experimental data are discussed next.\\\\
The close-ups in Figs.~\ref{SADAM}(h) and~\ref{NREX}(h) reveal that model 3 predicts a shift in the position of the 2nd order SLBP towards lower (higher) $q_z$ for the  $|-\rangle$ ( $|+\rangle$) curve. Such a shift is not observed in the experimental data where the maxima nearly coincide and the splitting of the curves is due to the different intensities of the peaks. The disagreement is especially obvious on the high $q_z$ side of the 2nd order SLBP where the intensity of the fitted $|+\rangle$ curve is higher than for the  corresponding $|-\rangle$ curve, whereas in the experimental data the opposite trend is observed. Such a discrepancy was already noted in Ref. \onlinecite{Stahn_2005} and was used to discard model 3. Furthermore, we note that a rather large value of the induced magnetization in the YBCO layers has to be assumed for model 3 to account for the large intensity of the 2nd order SLBP. For the fits in Figs~\ref{SADAM} and~\ref{NREX} the FM moment in the YBCO layer has been constrained to have the same value as the one in the LCMO layers. When it was released, the induced magnetization reached even larger values whereas the quality of the fit was not significantly improved. Already the constrained value appears to be unreasonably large, e.g. for the PNR curves taken at 4\,kOe it reaches $\sim 2.7\,\mB$. Such a large ferromagnetic moment on the YBCO side of the interface cannot arise from the induced Cu moments, it would also require an unrealistically large concentration of Mn ions.\\\\
For model 2, as shown in the close-ups in Figs~\ref{SADAM}(e) and~\ref{NREX}(e), can account reasonably well for the data in the vicinity of the 2nd order SLBP. Nevertheless, it largely overestimates the intensity of the 3rd order SLBP in the $|-\rangle$ curve. The intensity of the 3rd SLBP in the simulation could be reduced assuming an increased roughness of the magnetic potential. However, this would lead to a faster decay of the curve to the background level which is not observed. It would also further enhance the discrepancy at the 4th order SLBP where the fit already severely underestimates the peak intensity of the $|-\rangle$ curve. Furthermore, model 2 has the same problem as model 3 concerning the very large value of the induced moment in the YBCO layers that has to be assumed. Once more, for the fit in Figs~\ref{SADAM}(d) and~\ref{NREX}(d) the FM moment in the YBCO layer has been constrained to be the same as the one in the LCMO layers, i.e. at 4 kOe it reaches $\sim 2.7\,\mB$. This value is larger than the moment of $\sim 1\,\mB$ of Cu$^{+2}$ and one order of magnitude larger than the value reported from the XMCD measurements, $\sim 0.25\,\mB$~\cite{Satapathy_2012}. If the induced magnetic moment is for example bounded to $1\,\mB$, the intensity of the 2nd SLBP is largely reduced and the overall quality of the fit is strongly reduced.\\\\
Model 1 is clearly the one that reproduces the measured data the best and yields a realistic magnitude of the ferromagnetic moments. The position, spin splitting, and intensity of all SLBPs are reasonably well described. Only the overestimation of the intensity of the 4th order SLBP in the $|+\rangle$ curve at NREX can be regarded as a significant mismatch between the simulation and the data. The very fact that such a simple block-like model reproduces all features of the measurement is remarkable and confirms that it contains the main characteristics of the magnetic depth profile of the SL. These simulations clearly establish the trend that a sizable suppression of the FM moment on the LCMO side of the interface is responsible for the occurrence of the even order SLBPs. The characteristic length of the decay of the ferromagnetic moment at the interface should therefore be directly related to the calculated thickness of the depleted layers which are shown in Table~\ref{depl} for the different temperatures and fields.\\\\
It appears that the thickness of the depleted layer is consistently larger for the bottom interface than for the one on top of the LCMO layer (in terms of the SL growth direction). The origin of this difference between the two interfaces is not known. It may be related to a structural difference that is imposed by a different growth process. However, a recent transmission electron microscopy (TEM) study on superlattices that were grown under identical conditions did not show any clear indication of such a difference~\cite{Malik_2012}. Surprisingly, but in good agreement with a previous study of similar superlattices on STO substrates~\cite{Varela_2003}, the TEM images suggest that both the YBCO/LCMO and the LCMO/YBCO interfaces involve the same kind of CuO$_2$-Y-CuO$_2$-BaO-MnO$_2$ layer stacking sequence for which the last YBCO unit cell is lacking its CuO chains. The expected asymmetry of the interfaces, where a CuO chain layer should be adjacent to one of the interfaces and a CuO$_2$ bilayer to the other one, could not be observed. This still leaves the possibility that the layer separating the CuO$_2$ and MnO$_2$ planes may have a different stoichiometry, i.e. it may have a variable Ba and La or Ca content.\\\\
\begin{figure}[t!]
\centering
\includegraphics[width=8.6cm]{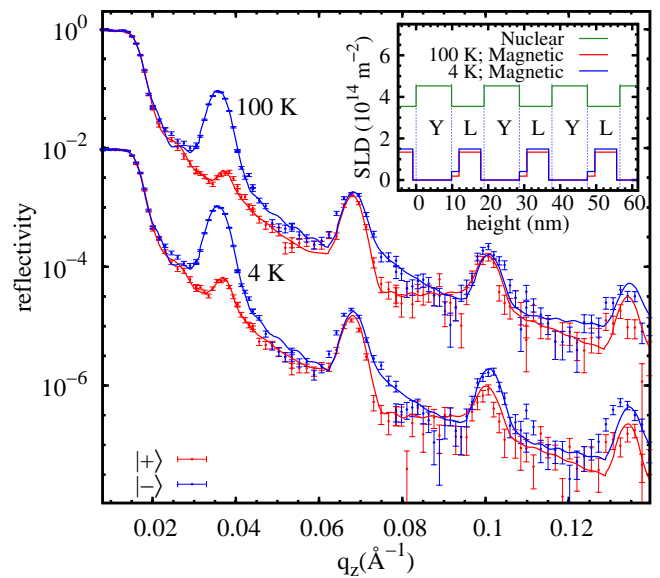}
\caption{Low temperature reflectivity curves measured under an applied field of 4\,kOe. The solid lines are the results of the fit using Model 1a, here the magnetization in the depleted layers is fitted and the thickness of the top and bottom interfaces are set as common among all datasets.\label{Model1a}}
\end{figure}
To obtain further information about the magnetic nature of the depleted layers, we have modified the fitting with model 1 by allowing for a finite magnitude of the FM moments in the depleted layer (treating the net magnetization in the depleted layers as a fitting parameter). To keep the number of fitting parameters reasonably low, we have now set the thickness of the depleted layers  for the top and bottom interfaces to be common for all temperatures and fields. This modification is labeled as model 1a and the comparison of the fitted curves using this model and the PNR data at 4\,kOe data is shown in Fig.~\ref{Model1a}. The inset shows the obtained depth profile of the scattering length densities, for which the magnetic part is proportional to the FM moment. In agreement with the results obtained for model 1, shown in Table~\ref{depl}, the thicknesses of the top and bottom depleted layers are calculated as 9.2(7)\,\AA\- and 20.1(7)\,\AA, respectively. It appears that the best fit obtained with model 1a is indeed very similar to the one of model 1~(see Fig.~\ref{NREX}) and results in the same value of the reduced likelihood estimator of $l^{\rm red}_{\rm 1a}=4.12$. The fit with model 1a yields a finite magnetic scattering length density near the bottom interface corresponding to a FM moment of about 20 percent of the one in the central part of the LCMO layers. This value increases somewhat as the temperature is reduced from 100\,K to 4\,K, in good agreement with model 1 where a corresponding decrease of the depleted layer thickness was obtained. Near the top interface the fit with model 1a does not yield a finite value of the FM moment. Nevertheless, given the crude assumption of block-like magnetic potentials, we are not sure whether these fits results are significant. Therefore we cannot draw any firm conclusion about presence of a small ferromagnetic moment in the depleted layers.\\\\
As a next step, we have addressed the question whether the induced ferromagnetic Cu moment in the YBCO layers are playing a significant role for the fitting of the PNR curves. The presence of these Cu moments was confirmed by the XMCD and XRMR measurements which suggest an average FM moment of 0.25\,$\mB$ per Cu ion that is antiparallel to the applied field and the moments in LCMO~\cite{Satapathy_2012}. We have therefore modified model 1 to allow in addition to the depleted layers on the LCMO side for a small, antiparallel moment on the YBCO side. It turned out that these Cu moments do not significantly modify or improve the fit results. This is not surprising since these Cu moments are about ten times smaller than the Mn moments in the central part of the LCMO layers and so are the corresponding value of the magnetic potential and the magnetic scattering length density. The reflectivity curves are  governed by the larger magnetic moments inside the LCMO layers and within the sensitivity of our PNR measurements we cannot make any quantitative statements regarding the induced Cu moment in the YBCO layers.
\begin{table}[t!]
\centering
\begin{tabular}{|c|cc|}\hline
$T,H$ & $d_{\rm{dep}}^{\rm{bottom}}$ (\AA) & $d_{\rm{dep}}^{\rm{top}}$ (\AA)  \\\hline\hline
10\,K, 100\,Oe & 14.0(7)  & 9.3(7)  \\\hline
100\,K, 100\,Oe & 17.9(7) & 8.5(7)   \\\hline
4\,K, 4\,kOe & 13.2(7) & 11.0(7) \\\hline
100\,K, 4\,kOe & 19.0(7) & 8.4(7) \\\hline
\end{tabular}
\caption{Thickness of the depleted layers at the bottom and top interfaces as obtained with model 1 at different temperatures and applied magnetic fields.\label{depl}}
\end{table}\\\\
Finally, we have tested the validity of the fits as obtained with models 1 - 3 by comparing the average magnetic moment of the fitted profile (see Figs.~\ref{SADAM} and~\ref{NREX} c, f, and i) with the experimental value as obtained from dc magnetization measurements. The result is summarized in Fig.~\ref{FC} which shows the magnetic moment from the field-cooled dc magnetization measurements at 100\,Oe and 4\,kOe (solid lines) together with the magnetic moments as calculated from the best fits of models 1, 1a, 2 and 3 to the PNR curves  at 100\,Oe and 4\,kOe (symbols). This comparison shows a very good agreement for models 1 and 1a and large discrepancies for models 2 and 3. For model 3 where the FM moments in the LCMO and YBCO layers are assumed to be parallel, the calculated magnetic moment is almost twice as large as the measured one. For model 2 the calculated magnetic moments are significantly smaller than the measured ones. Evidently, this is the consequence of the assumption that the FM moment in the YBCO layers is antiparallel to the ones in the LCMO layers. This comparison clearly argues against models 2 and 3 and emphasizes the relevance of models 1 and 1a in terms of the depleted FM layer on the LCMO side of the interfaces. The discrepancy between the calculated magnetic moment of models 1 and 1a and the experimentally measured value never exceeds 15\% and it is almost within the error bars. Given the simplicity of the model, with its simple block like potentials, this agreement can be considered as excellent. The somewhat larger difference that appears in the low field data at 100\,Oe, where the magnetization is not yet fully saturated, may have its origin in a weak spin flip scattering which would contribute to both spin channels, especially around the 1st Bragg peak, and thus enhance the calculated magnetic moment.
\\\\
To conclude, the analysis of the PNR data as shown above provides clear evidence for the presence of a so-called depleted layer on the LCMO side of the interfaces in which the ferromagnetic moment of the Mn ions is strongly suppressed as compared to the central part of the LCMO layers. These depleted layers have a sizable thickness and therefore are not likely just the result of chemical and/or structural disorder. At the top interface the depleted layer extends over about 3 LCMO unit cells (with a lattice parameter of $~3.9$\,\AA) and at the bottom interface it involves even 4-5 LCMO unit cells. The depleted layer thickness also has an unusual temperature dependence. At the top interfaces it remains almost constant, whereas at the bottom interfaces it decreases considerably toward low temperature. As reported in Ref.~\onlinecite{Satapathy_2012}, this decrease is even anomalously enhanced below the superconducting transition temperature. Additional evidence for an intrinsic electronic origin of the depleted layers in the YBCO/LCMO SL comes from the observation that the magnetic proximity effect and thus these depleted layers are absent for a corresponding YBCO/LaMnO$_{3+\delta}$ SL for which the manganite layers are insulating~\cite{Satapathy_2012}. The large FM moments of the Mn ions persist here right up to the interface. Furthermore, there is the observation of the XMCD and XRMR studies that a ferromagnetic (or strongly canted antiferromagnetic) order of the Cu moments is induced on the YBCO side of the interfaces of the YBCO/LCMO SL. Notably, these Cu moments are much weaker or even absent in the YBCO/LMO SLs where the FM order of the Mn moments persists right up to the interface. These observations suggest that the induced ferromagnetic Cu moments in the YBCO layers are a central part of the magnetic proximity effect just as much as the suppression of the FM moment on the LCMO side of the interface. The apparent antiparallel coupling between the induced Cu moments in the YBCO layers and the Mn moments therefore requires that the depleted layer maintains some kind of magnetic order. Likely, this involves a non-collinear magnetic order that cannot be detected with the PNR technique which probes the average FM component.\\\\
Finally, we note that such a non-collinear magnetic order may have important consequences for the superconducting proximity effect in these YBCO/LCMO superlattices. It was previously shown that it can induce a spin-triplet component of the superconducting order parameter which has a long-ranged proximity effect into the FM layers~\cite{Eschrig_2011,Bergeret_2005}. Evidence for such as scenario has indeed been reported very recently based on the observation of equal-spin Andreev reflections in YBCO/LCMO interfaces~\cite{Visani_2012}.
\section{Summary}
We have performed wide $q_z$ range PNR measurements in a YBCO/LCMO SL for different applied fields and temperatures. The emergence of the 2nd order SLBP below $\tcur$ evidences a mismatch of the magnetic potential with respect to the nuclear one. After fitting the data with three different models and comparing the results we have ruled out the possibility of an induced magnetic moment in YBCO as the main origin of the asymmetry. Our results suggest that the asymmetry in the potential mainly comes from the existence a so-called depleted layer in the LCMO side of the interface where the net FM moment is strongly reduced. The characteristic length of this reduction being of $\sim 1$\,nm is larger than the expected interface roughness suggesting an electronic, rather than an structural origin for the depletion zones. The actual magnetic state of the depleted layers could consist of some canted or oscillatory order as suggested by the induction of a Cu moment in the YBCO side and the observation of long--range spin--triplet correlations inside the FM layers in a similar system.
\acknowledgements
The UniFr group was supported by the SNF Grants No. 200020-129484 and 200020-140225 and the NCCR MaNEP. We acknowledge financial support from the European Commission under the 7th Framework Programme through the \textquoteleft Research Infrastructures\textquoteright\ action of the \textquoteleft Capacities\textquoteright\ Programme, NMI3-II Grant number 283883; and the Deutsche Forschungsgemeinschaft within the framework of the TRR80, project C1. This work is based on experiments performed  NREX at FRM-II, Munich, D, and on SuperADAM at ILL, Grenoble, F.

%


\end{document}